\documentclass[11pt,a4paper]{article}
\usepackage[T1]{fontenc}
\usepackage[utf8]{inputenc}
\usepackage{lmodern}
\usepackage[english]{babel}
\usepackage{amsmath,amssymb}
\usepackage{graphicx}
\usepackage[margin=2.5cm]{geometry}
\usepackage[font=small,labelfont=bf]{caption}
\usepackage{xcolor}
\usepackage[colorlinks=true,linkcolor=blue!50!black,citecolor=blue!50!black,urlcolor=blue!50!black]{hyperref}
\renewcommand{\d}{\mathrm{d}}
\newcommand{\e}{\mathrm{e}}
\newif\ifmarked \markedfalse
\newcommand{\rev}[1]{\ifmarked{\color{red}#1}\else#1\fi}
\newcommand{\revold}[1]{#1}% revisions of v1, now accepted

\title{\textbf{Oblique incidence and non-commuting limits:\\
a remark on the convergence of Fresnel integrals}}
\author{\rev{Giancarlo Reali}\\[2pt]
\small \rev{Bright Solutions S.r.l., Via degli Artigiani 27, 27010 Cura Carpignano (PV), Italy}\\
\small \rev{\texttt{g.reali@brightsolutions.it} \quad -- \quad ORCID 0000-0001-9943-5079}}
\date{}

\begin{document}
\maketitle

\begin{abstract}
\noindent The Gaussian integrals of quantum mechanics and of paraxial optics are oscillatory and
only conditionally convergent. Textbooks usually make them converge by adding a small imaginary
part to the time or by analytic continuation. We point out that the one-dimensional Fresnel
integral converges by itself, while in two dimensions the result depends on the shape of the
expanding domain: over squares it converges, over discs it oscillates for ever. The second case is
the field of a uniformly excited plane of dipoles, which Feynman uses to explain the refractive
index. Lai, Lau and Wong showed that at oblique incidence the integral converges and that normal
incidence is recovered in the limit; we note that the two limits do not commute, and that this
provides a transparent introduction to the regularization of the time-sliced path integral, where
the ordering of the integrations in time plays the role of the physical prescription.
\end{abstract}

\section{Introduction}

The free propagator of quantum mechanics and the paraxial propagator of optics have the same
structure, $\propto\exp[i\,a\,(x_b-x_a)^2/2]$, and both are composed by integrals of the type
\begin{equation}
\int_{-\infty}^{\infty}\e^{\,iax^2/2}\,\d x=\Big(\frac{2\pi}{|a|}\Big)^{1/2}\e^{\,i\frac{\pi}{4}\operatorname{sgn}a}.
\label{eq:fresnel}
\end{equation}
The integrand has modulus one, and the integral is not absolutely convergent. In quantum mechanics
it is usually made convergent by giving the time a small negative imaginary part, or by analytic
continuation from imaginary time;\cite{schulman,feynmanhibbs} in the time-sliced path integral one
meets $N-1$ such integrals at once.\cite{schulman} The prescription works, but it hides two facts
that a simple problem of classical optics makes visible: in one dimension the integral needs no
help, and in more dimensions it is not defined until a prescription is chosen. \revold{The damping is
not wrong in one dimension, only unnecessary; it becomes indispensable in more.}

\section{One dimension: the integral converges by itself}

With $u=x^2$,
\[
\int_0^R\e^{iax^2/2}\,\d x=\frac12\int_0^{R^2}\e^{iau/2}\,u^{-1/2}\,\d u ,
\]
and since $u^{-1/2}$ decreases to zero the improper integral converges (Dirichlet's test). The
remainder beyond $R$ is of order $1/(aR)$, because the local frequency of the oscillations grows
linearly with $x$: the integrand is a ``chirp''. In the complex plane the partial integral traces
the Cornu spiral [Fig.~\ref{fig:conv}(a)], which winds into the limit point
$(2\pi/a)^{1/2}\e^{i\pi/4}$. In optical language, the Fresnel zones of a slit have widths
decreasing\revold{, for large $n$,} as $n^{-1/2}$, and their alternating contributions form a
convergent series.\revold{\cite{bornwolf}} No damping is needed.

\section{Two dimensions: the shape of the domain matters}

Consider now
\begin{equation}
I=\int\!\!\int\exp\Big[\frac{i}{2}(x^2+y^2)\Big]\d x\,\d y ,
\label{eq:twod}
\end{equation}
which, apart from constants, is the paraxial field on the axis at distance $z$ from a plane of
dipoles uniformly excited by a wave at normal incidence: the problem that Feynman uses to explain
the origin of the refractive index,\cite{feynmanlectures} \rev{and that is developed step by step in Ref.~\cite{lightmatter}.} \revold{Feynman works with the exact
spherical waves, for which the integral reduces to $\int\e^{ikr}\,\d r$ (see below); Eq.~(\ref{eq:twod})
is its paraxial form, and has the same difficulty.} For any domain bounded by a curve
$\rho=\rho_{\max}(\varphi)$\revold{ (a domain star-shaped about the origin)}, integrating first along
the radius\revold{, with $\int_0^{\rho_{\max}}\e^{i\rho^2/2}\rho\,\d\rho=i\big(1-\e^{i\rho_{\max}^2/2}\big)$,}
gives
\begin{equation}
\int_{\rm domain}\e^{i\rho^2/2}\,\d A=2\pi i-i\int_0^{2\pi}\e^{\,i\rho_{\max}(\varphi)^2/2}\,\d\varphi .
\label{eq:boundary}
\end{equation}
The first term is the \revold{regularized} value; the second is the contribution of the boundary.
Over squares $|x|,|y|<R$, $\rho_{\max}^2$ varies by a quantity of order $R^2$ along the boundary, the
angular integral is dominated by a few stationary points and decreases as $1/R$, and the integral
converges to $2\pi i$ (it is the square of the one-dimensional integral). {\relax The same holds
whenever $\rho_{\max}^2(\varphi)$ has isolated, non-degenerate stationary points, as for an ellipse;
degenerate stationary points give a slower decay, but still a decay. The boundary term fails to
vanish only if $\rho_{\max}$ is constant over some interval of angles, i.e.\ if part of the boundary
is an arc of a circle centred on the origin. The extreme case is the disc.} For $\rho_{\max}=R$,
\begin{equation}
I(R)=2\pi i\big(1-\e^{iR^2/2}\big),
\end{equation}
which oscillates for ever around the correct value [Fig.~\ref{fig:conv}(b)]. On a circle the phase
is the same at all points of the boundary, and its contribution does not decay; in the language of
Fresnel zones, the circular zones centred on the axis have equal areas, and their alternating
contributions do not decrease. Feynman disposes of the oscillating term by an argument on the
contribution of the very distant charges,\cite{feynmanlectures} \revold{which amounts to letting
their contributions fade out gradually: a soft aperture (Sec.~5)\revold{; a careful version of this argument, with an exponential taper of
the density and the order of the limits made explicit, is given by Galiautdinov\cite{galiautdinov}}. Nor is the difficulty an artefact of
the paraxial approximation: with the exact spherical wave, $\rho\,\d\rho=r\,\d r$ and the field on
the axis is $2\pi\int_z^{r_{\max}}\e^{ikr}\d r$, whose boundary term $\e^{ikr_{\max}}$ does not
decay either\revold{; its phase grows linearly with $r_{\max}$ rather than quadratically, but the
contrast between disc and square is the same}.}

\begin{figure}[t]
\centering
\includegraphics[width=\textwidth]{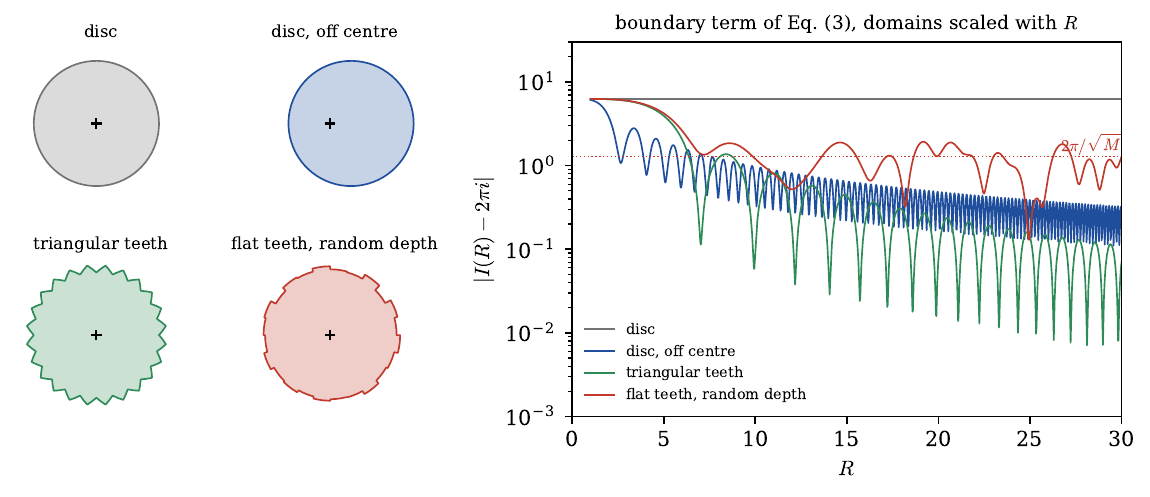}
\caption{\revold{Boundary term $|I(R)-2\pi i|$ of Eq.~(\ref{eq:boundary}) (absolute value, not
normalized) for four domains, all scaled with $R$ (the cross marks the axis). Disc: constant, equal to $2\pi$. Disc displaced by $R/3$:
decays as $1/R$. Disc with $M=24$ triangular teeth of depth $0.12R$: decays. Disc with $M=24$
flat-topped teeth of random depth (up to $0.12R$): stays of order $2\pi/\sqrt M$; the dotted line is this estimate for $M$ terms of random phase,
not a fit.}}
\label{fig:shapes}
\end{figure}

{\relax
Figure~\ref{fig:shapes} shows what it takes to break the coherence of the boundary. Moving the disc
off the axis is enough: $\rho_{\max}$ now varies along the boundary, and the boundary term decays as
$1/R$. Asymmetric surfaces are used for this purpose by Hopf and Stegeman.\cite{hopfstegeman}
Serrating the edge works too, provided the teeth are slanted, as in the triangular teeth of the
figure: within each tooth $\rho_{\max}$ varies, and the contributions no longer add in phase. Teeth
with flat tops, instead, are arcs of circles centred on the axis, and each of them contributes a
term that does not decay. Giving the $M$ teeth random depths only randomizes the phases of these
terms: their sum is reduced to about $2\pi/\sqrt M$, but does not vanish. On the axis only the
distribution of the boundary radii matters, not the arrangement of the teeth: the boundary term in
Eq.~(\ref{eq:boundary}) is $\int\e^{i\rho^2/2}g(\rho)\,\d\rho$, where $g(\rho)\,\d\rho$ is the angle
over which the boundary lies between $\rho$ and $\rho+\d\rho$. A serrated edge is thus equivalent,
on the axis, to a soft aperture (Sec.~5) whose transmission at radius $\rho$ is the fraction of the
angle still open there; irregular teeth act off the axis: a periodic serration is an angular
grating, which sends light into discrete off-axis directions, and irregular teeth spread it. The same idea, breaking
the phase coherence of a hard edge, underlies the serrated apertures used in high-power lasers to
suppress the diffraction ripples of the beam.\cite{auerbach}
\par}

\section{Oblique incidence and a non-commuting double limit}

\begin{figure}[t]
\centering
\includegraphics[width=\textwidth]{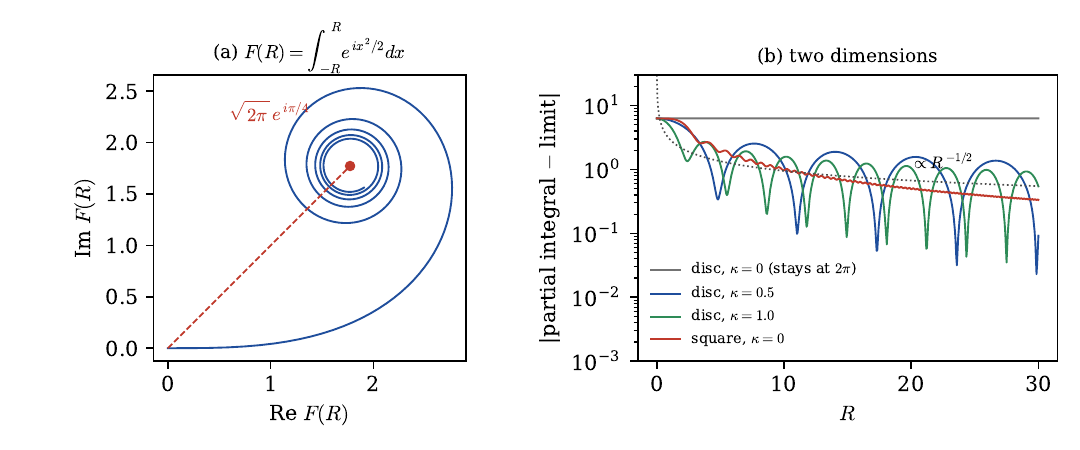}
\caption{(a) The partial Fresnel integral $F(R)=\int_{-R}^{R}\e^{ix^2/2}\d x$ traces the Cornu
spiral and converges to $\sqrt{2\pi}\,\e^{i\pi/4}$ without any damping. (b) Distance of the
two-dimensional partial integrals of $\exp[i(x^2+y^2)/2+i\kappa x]$ from their limits
$2\pi i\,\e^{-i\kappa^2/2}$. Over squares the integral converges. Over discs at normal incidence
($\kappa=0$) the distance stays equal to $2\pi$: \revold{the partial integral runs for ever around a
circle of radius $2\pi$ centred on the limit}. At oblique incidence ($\kappa=0.5$, $1$) the
distance decreases as $R^{-1/2}$ (dotted line), and the limit $\kappa\to0$ can be taken afterwards.
\revold{Ranges: $0\le R\le8$ in (a), $0<R\le30$ in (b).}}
\label{fig:conv}
\end{figure}

Lai, Lau and Wong\cite{llw} showed that the difficulty disappears if the wave is incident
obliquely. The excitation of the plane then carries the phase $\e^{i\kappa x}$, with $\kappa$
proportional to the sine of the angle of incidence, and
\begin{equation}
I(R,\kappa)=\int_{\rho<R}\e^{\,i(\rho^2/2+\kappa x)}\,\d A
=2\pi\int_0^RJ_0(\kappa\rho)\,\e^{i\rho^2/2}\rho\,\d\rho
\;\xrightarrow[R\to\infty]{}\;2\pi i\,\e^{-i\kappa^2/2}.
\label{eq:oblique}
\end{equation}
Along the circle the phase $\kappa R\cos\varphi$ now varies, the angular average
$J_0(\kappa R)\sim(\kappa R)^{-1/2}$ kills the boundary term, and the integral converges
\revold{as} $R^{-1/2}$ [Fig.~\ref{fig:conv}(b)]\revold{; this is the case $d=2$ of the estimate of
Sec.~5}. Normal incidence is then obtained as the limit $\kappa\to0$ of
the converged result. The two limits do not commute:
\begin{equation}
\lim_{\kappa\to0}\,\lim_{R\to\infty}I(R,\kappa)=2\pi i,\qquad
\lim_{R\to\infty}\,\lim_{\kappa\to0}I(R,\kappa)=\lim_{R\to\infty}2\pi i\big(1-\e^{iR^2/2}\big),
\end{equation}
\revold{and the second limit does not exist.}
The physically reasonable order is the first one: a real wave is never exactly at normal incidence,
a real plane is never exactly infinite, a real medium is never exactly lossless. The example shows
to students, better than any general statement, that a conditionally convergent multiple integral
is defined only together with a prescription, and that physics selects the prescription.

\section{Path integrals}

The time-sliced propagator of quantum mechanics is a Fresnel integral in $N-1$ dimensions, with
$N\to\infty$.\cite{schulman,feynmanhibbs} The oblique-incidence cure does not carry over: a rough
estimate of the boundary term over a ball of radius $R$ in $d$ dimensions, with a linear phase,
gives $R^{d-1}$ (area) $\times(\kappa R)^{-(d-1)/2}$ (angular average) $\times R^{-1}$ (inverse
radial gradient of the phase), i.e.\ $R^{(d-3)/2}$, which \revold{among integer dimensions} decreases only for $d=2$. \revold{The
estimate is heuristic, and assumes $\kappa R\gg1$ and $R\gg\kappa$; it concerns this particular
family of domains, not every regularization, but its exponent is confirmed numerically below, also
for non-integer $d$. In $d=3$ it can be checked exactly: the angular average is
$\sin(\kappa\rho)/\kappa\rho$, and the partial integral keeps oscillating with an amplitude of
order $4\pi/\kappa$.} \revold{Three} prescriptions do work in any dimension.
\begin{enumerate}\itemsep2pt
\item \emph{A soft aperture.} Multiply the integrand by a smooth cut-off, for instance
$\e^{-\delta|\mathbf x|^2}$, and let $\delta\to0$. For a Gaussian cut-off this is the same as the
imaginary part added to the time\revold{: with $t\to t(1-i\epsilon)$ the exponent $iax^2/2$ of the
free propagator acquires the damping $-\epsilon ax^2/2$, i.e.\ $\delta=\epsilon a/2$}. It is also the definition of an oscillatory integral used in the
rigorous theory of Feynman integrals, due to Hörmander:\cite{hormander,ahk} the limit is required to
exist and to be the same for every smooth cut-off $\phi(\delta\mathbf x)$ with $\phi(0)=1$. The
requirement of independence from the cut-off is the precise answer to the disc-versus-square
ambiguity: a disc with a sharp edge is not an admissible cut-off\revold{, its indicator function
being discontinuous. Neither is a square; that it converges anyway is a fortunate consequence of
its factorization into one-dimensional integrals. The criterion is sufficient, not necessary.}
\item \emph{The order of time.} Integrate the intermediate positions $x_1,x_2,\dots,x_{N-1}$ one at a
time, in the order of time, completing the square at each step. Each step is a one-dimensional
Fresnel integral, which converges by itself, and the result of each step is again of the same
form. For the quadratic actions of semiclassical physics this iterated integral needs no
regularization at all, and the ordering is not arbitrary: it is the composition of short-time
propagators, one time slice after the other. {\relax The distinction is the one of Sec.~4: the
$(N-1)$-dimensional integral taken as a whole, over growing balls, is not defined; the iterated
integral is, because the order of the limits is part of its definition.} The existence of the limit
$N\to\infty$ for \revold{a broad class of smooth potentials} has been proved by D.~Fujiwara and by
Kumano-go, with the time-sliced integrals understood as oscillatory
integrals.\cite{dfujiwara,kumanogo,ahk}
{\relax
\item \emph{Continuation in the dimension.} The radial reduction of Eq.~(\ref{eq:oblique}) holds in any dimension,
integer or not. Averaging $\e^{i\boldsymbol\kappa\cdot\mathbf x}$ over the sphere,
\begin{equation}
I_d(R,\kappa)=(2\pi)^{d/2}\kappa^{1-d/2}\!\int_0^R\!\rho^{d/2}J_{d/2-1}(\kappa\rho)\,
\e^{i\rho^2/2}\d\rho
\;\xrightarrow[R\to\infty]{}\;(2\pi i)^{d/2}\e^{-i\kappa^2/2},
\label{eq:dimd}
\end{equation}
and for $\kappa=0$, $I_d(R)=[2\pi^{d/2}/\Gamma(d/2)]\int_0^R\rho^{d-1}\e^{i\rho^2/2}\d\rho$.
Computed numerically \revold{(with $\kappa=1$, for $d=2$, $2.5$, $2.8$, $2.9$, $3$, $3.1$, $3.2$,
$3.5$, taking as envelope the maximum distance from the limit over the windows $15<R<20$ and
$50<R<60$; script \revold{among the ancillary files of this preprint})}, the oscillation of $I_d(R,\kappa)$ about its
limit has an envelope that follows $R^{(d-3)/2}$ \revold{to within $0.01$ in the exponent}: the oblique-incidence cure works in
$3-\varepsilon$ dimensions, slowly, and fails in $3+\varepsilon$. At normal incidence the disc
integral converges for $d<2$, as $R^{d-2}$, to $(2\pi i)^{d/2}$. The limit is an analytic function
of $d$; computed where the integral converges and continued analytically, it defines the integral
in any dimension, including the $N-1$ of the path integral, and even the disc at normal incidence
is no longer singular when seen from $d<2$. This is the dimensional regularization of quantum
field theory\revold{\cite{thooft}---the analogy concerns the regularization step only, not the
renormalization that follows it in that context}. A non-integer dimension has no counterpart on the optical bench: here
the prescription is justified not by the correspondence between model and apparatus, but by the
correspondence between model and result. The three prescriptions are of different nature,
experimental, dynamical and formal, and give the same value wherever more than one of them applies.
\par}
\end{enumerate}

\section{Conclusion}

The one-dimensional Fresnel integral converges without damping; the two-dimensional one depends
on the shape of the domain, and the disc, which corresponds to normal incidence on an infinite
plane, is the one shape for which it does not converge. Lai, Lau and Wong's oblique incidence
removes the difficulty at the price of two non-commuting limits. In the path integral, where the
dimension is unbounded, the same role is played by a smooth aperture, by the
physically natural ordering of the integrations in time\revold{, or, formally, by continuation in the
dimension}. \revold{A conditionally convergent multiple
integral is not defined by its integrand alone: it is a limit, and physics chooses the limit.}

\section*{Acknowledgements}
\revold{The author used Claude (Anthropic), an AI assistant, to check calculations, run the numerical
examples and edit the text; he reviewed all the material and takes full responsibility for the
content.}

\end{document}